\documentclass[twocolumn]{aastex631}
\usepackage[T1]{fontenc}
\usepackage{amssymb}
\usepackage{amsmath}
\usepackage{xspace}
\usepackage{upgreek}
\usepackage{graphbox}
\usepackage{graphicx,xparse,booktabs}

\usepackage{hyperref}
\usepackage{graphicx}
\usepackage{amsmath,amssymb}
\usepackage{subfigure}
\usepackage{rviewport}
\usepackage{xspace}
\usepackage{xcolor}
\usepackage{multirow}
\usepackage{mathptmx}

\DeclareMathAlphabet{\mathcal}{OMS}{cmsy}{m}{n}

\newcommand{\R}{\ensuremath{{\mathcal R}\xspace}}
\newcommand{\atEarth}{\scalebox{0.5}{$\oplus$}}

\definecolor{rossoCP3}{cmyk}{0,.88,.77,.40}
\definecolor{darkBlue}{rgb}{0, 0, 0.8}
\hypersetup{
    bookmarks=true,         
    bookmarksopen=true,     
    bookmarksopenlevel=1,
    unicode=false,          
    pdftoolbar=true,        
    pdfmenubar=true,        
    colorlinks=true,        
    linkcolor=darkBlue,     
    citecolor=darkBlue,     
    filecolor=darkBlue,     
    urlcolor=darkBlue       
}

\begin{document}

\title{Where Did the Amaterasu Particle Come From?}

\author[0000-0002-7651-0272]{Michael Unger}
\email{michael.unger@kit.edu}
\affiliation{Institut f\"ur Astroteilchenphysik, Karlsruher Institut f\"ur Technologie, Karlsruhe 76344, Germany}
\affiliation{Institutt for fysikk, Norwegian University of Science and Technology (NTNU), Trondheim, Norway}
\author[0000-0003-2417-5975]{Glennys R. Farrar }
\email{gf25@nyu.edu}
\affiliation{Center for Cosmology and Particle Physics, Department of Physics, New York University, NY 10003, USA
}

\begin{abstract}
  \noindent The Telescope Array Collaboration recently reported the
  detection of a cosmic-ray particle, ``Amaterasu," with an extremely
  high energy of $2.4\times10^{20}$~eV.  Here we investigate its
  probable charge and the locus of its production.  Interpreted as a
  primary iron nucleus or slightly stripped fragment, the event fits
  well within the existing paradigm for UHECR composition and
  spectrum. Using the most up-to-date modeling of the Galactic
  magnetic field strength and structure, and taking into account
  uncertainties, we identify the likely volume from which it
  originated.  We estimate a localization uncertainty on the source
  direction of 6.6\% of $4\uppi$ or 2726 deg$^2$. The uncertainty of
  magnetic deflections and the experimental energy uncertainties
  contribute about equally to the localization uncertainty.  The
  maximum source distance is 8-50 Mpc, with the range reflecting the
  uncertainty on the energy assignment.  We provide sky maps showing
  the localization region of the event and superimpose the location of
  galaxies of different types.  There are no candidate sources among
  powerful radio galaxies. An origin in active galactic nuclei or
  star-forming galaxies is unlikely but cannot be completely ruled out
  without a more precise energy determination. The most
  straightforward option is that Amaterasu was created in a transient
  event in an otherwise undistinguished galaxy.
\end{abstract}

\section{Introduction}
Recently, the Telescope Array (TA) Collaboration reported
the detection of an air shower initiated
by a cosmic-ray particle with an estimated energy of $E =\left(2.44 {\pm}
0.29\, (\text{stat.}) \, ^{+0.51}_{-0.76} \, (\text{syst.})\right)
\times 10^{20}$~eV~\citep{TAScience}. The arrival direction was
$(\text{RA},\text{Dec}) = (255.9{\pm}0.6,\, 16.1{\pm}0.5)^\circ$ in
equatorial coordinates, or $(\ell, b) = (36.2,\,
30.9)^\circ$ in Galactic coordinates.  The TA Collaboration has named
the event ``Amaterasu."

At the nominal reconstructed energy, the Amaterasu event is the
second most energetic particle ever recorded after the famous Fly's
Eye event ($E=\left(3.2 {\pm} 0.9\, (\text{tot.})\right) \times
10^{20}$~eV)~\citep{HIRES:1994ijd}. Two other extremely energetic
events, within one standard deviation (total energy uncertainty)
of the Amaterasu particle, have been previously reported by the Pierre
Auger Collaboration~\citep{PierreAuger:2022axr,PierreAuger:2022qcg}.

Our purpose in this Letter is to characterize the distance and direction
of Amaterasu's source, based on its reported energy and what is known
about ultrahigh-energy cosmic rays (UHECRs) in general from earlier
measurements of their energy spectrum and mass composition.

Assuming that the air shower was initiated by an iron nucleus (see
the discussion in the next section), the nominal energy estimate of the
Amaterasu event is $E_\text{nom} = (2.12{\pm}0.25) \times 10^{20}$~eV,
including the quoted corrections for resolution effects ($-3\%$) and
heavy primaries ($-10\%$). Taking into account the 20\% systematic
energy scale uncertainty of Telescope Array, a conservative energy
estimate is $E_\text{low} = (1.64{\pm}0.19) \times 10^{20}$~eV. The
quoted uncertainty in both cases is due to the statistical error on
the measurement of the particle density at ground level.

We approach the analysis in steps.  In Sec.~\ref{sec:rigidity} we
investigate the joint probability distribution in source distance and
UHECR rigidity upon reaching the Galaxy, for each of the energy
assignments, $E_\text{nom}$ and $E_\text{low}$, based on the hypothesis that the original UHECR was an
iron nucleus.  With this information, in Sec.~\ref{sec:dist} we find
the probability distribution of source distance.  For each rigidity
value, in Sec.~\ref{sec:arrdir} we backtrack the arriving UHECR
through the Galactic magnetic field (GMF) to find the source direction
taking into account the uncertainties due to deflections of the coherent
and turbulent GMF. Integrating over the
rigidity distribution, we thereby find the localization uncertainty of the
particle.  Thus prepared, in Sec.~\ref{sec:cata} we investigate the
galaxies falling in the source volume; we conclude in
Sec.~\ref{sec:concl}.

\begin{figure*}[t]
  \begin{minipage}{0.46\textwidth}
  \includegraphics[width=\textwidth]{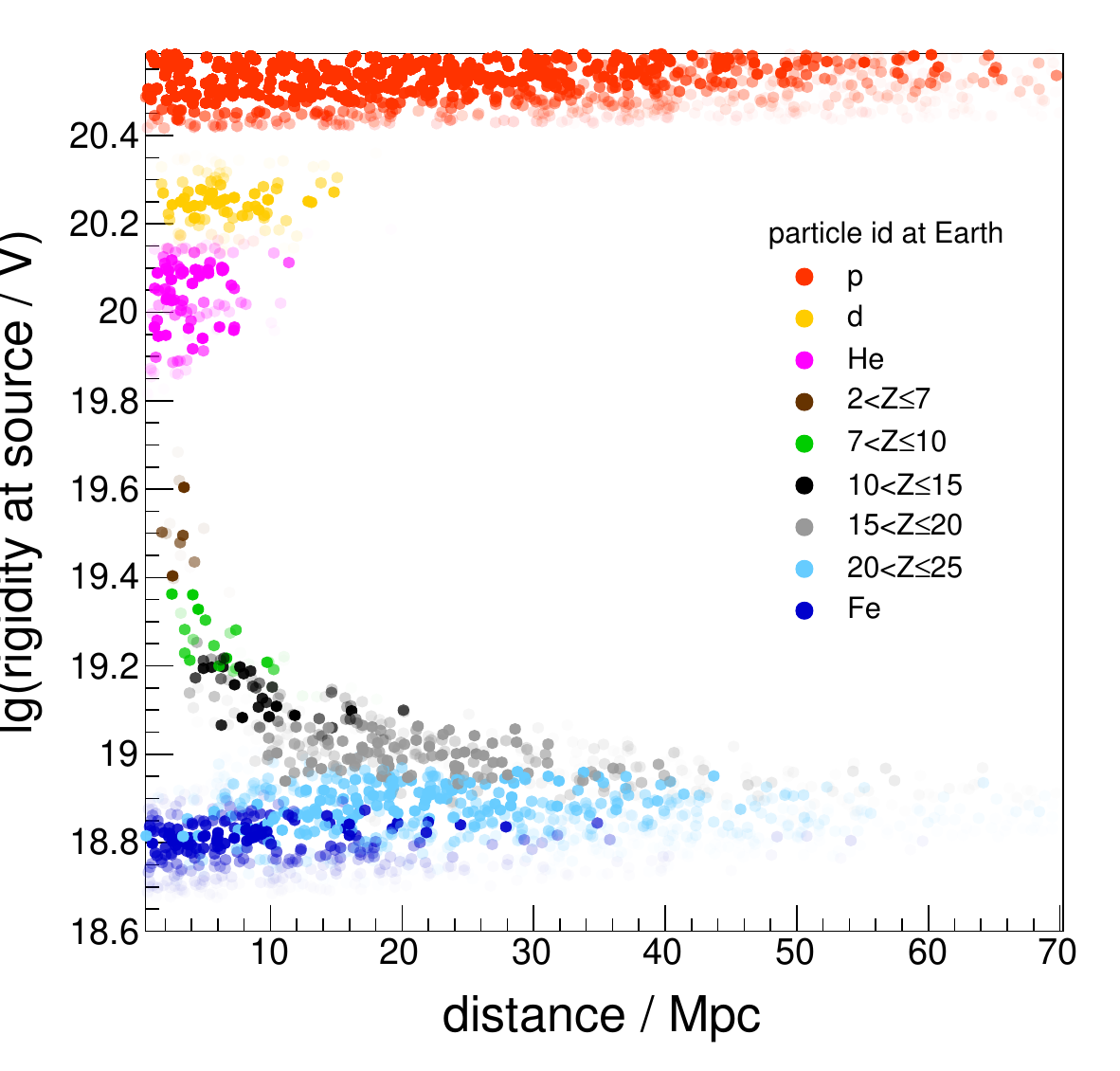}
  \end{minipage}\hfill
  \begin{minipage}{0.46\textwidth}
    \includegraphics[width=\textwidth]{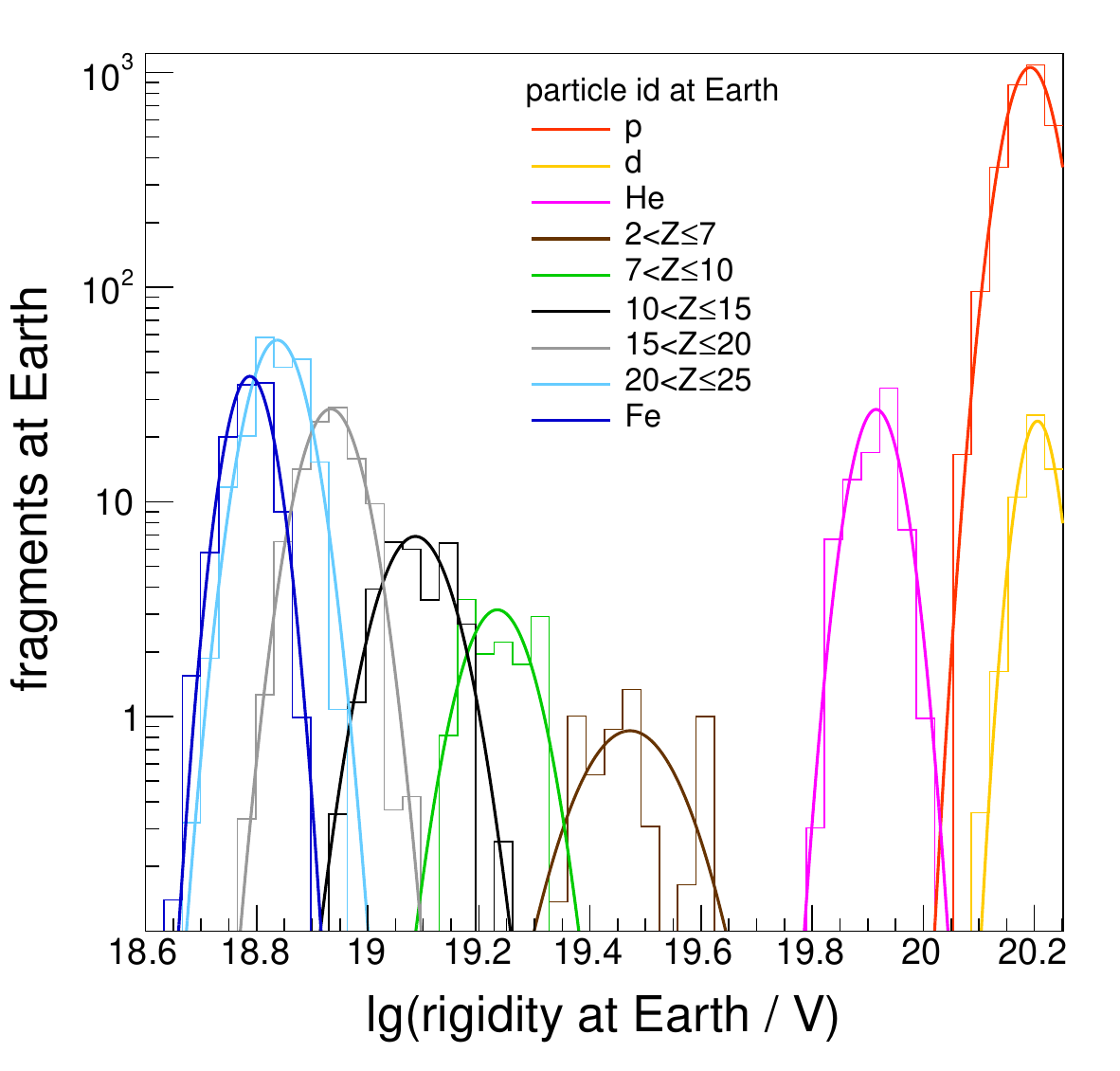}
  \end{minipage}
  \caption{Simulation of the propagation through extragalactic photon
    fields of a UHECR which originates as $^{56}_{26}$Fe, with a
    uniform distance distribution and an $E^{-1}$ energy spectrum (no
    cutoff) at the source. {\itshape Left:} Injected rigidity
    vs.\ source distance. Each point denotes a fragment of an iron
    nucleus that reaches Earth with an energy similar to the one of
    the Amaterasu particle. The transparency of each point encodes the
    likelihood the energy is reconstructed to be $E_\text{low}$,
    $w_E$.  The particle type of the fragments arriving at Earth are
    shown as colors. {\itshape Right:} Rigidity distribution of the
    fragments arriving at Earth. Lines are Gaussian fits to the
    histograms. Histogram entries are weighted according to their
    deviation from the measured energy of the TA event, $w_E$.  These
    plots are for the low energy scale for the Amaterasu
    particle.\label{fig:rig}}
\end{figure*}

\section{Particle Rigidity}
\label{sec:rigidity}

Magnetic interactions of relativistic charged particles depend on
their rigidity, and to good accuracy, the heavy fragments of
photo-dissociation of nuclei have the same rigidity as the
parent.  This motivates the ``Peters cycle''~\citep{Peters:1961}
ansatz in which all nuclei except spallation protons have a common
spectrum when expressed in terms of rigidity, only differing in their
normalization. A standard choice is a power-law rigidity spectrum with
an exponential cutoff, $\R^{-\gamma} \exp(-\R/\R_\text{max})$.

An excellent description of the observed spectrum and composition of
extragalactic cosmic rays at ultra-high energies can be obtained if a
maximum rigidity in the range of $10^{18.5}< \R_\text{max}/\text{V} <
10^{18.7}$ is assumed, under a wide range of assumptions including a
pure Peters cycle or modified scenarios with photonuclear or hadronic
interactions in the source environment \citep[see, e.g.,][,
  respectively]{PierreAuger:2016use, Unger:2015laa, Muzio:2021zud}.
Given $\R_\text{max}$ in this range, the accelerators are powerful
enough to produce an Amaterasu particle in the tail of the rigidity
spectrum if it is an iron nucleus ($Z=26$), since,
e.g.,\ $\exp(-E_\text{nom}/(26\times 10^{18.7}\text{V}))=
0.15$\footnote{A subdominant population of more powerful accelerators
  might exist with a best-fit maximum rigidity in the range of
  $10^{19.1}< \R_\text{max}^\text{sub}/\text{V} <
  10^{19.5}$~\cite{Muzio:2019leu,PierreAuger:2022atd,Ehlert:2023btz}. However,
  Amaterasu has too high an energy for it to be identified as a part
  of this subdominant proton population.}.

For the air shower detected by TA to have been initiated by a
proton, some entirely new acceleration mechanism must be invoked.  On
the other hand,
since the event fits well to the ultrahigh-energy end of the energy
spectrum reported by~\citet{Kim:2023eul} and since the TA flux at
these energies is well described by iron nuclei \citep[see Fig.~8
  in][]{Unger:2015laa}, the minimal assumption on the nature of the
particle is that it is an iron nucleus from the bulk of the cosmic-ray
flux at UHE. Therefore, we focus on iron nuclei injected at the source as the most
plausible origin of Amaterasu.

We next constrain the distance and rigidity of the initial iron cosmic
ray which evolved into the observed Amaterasu particle detected at
Earth.  The source distance will limit what extragalactic objects are
candidates to be the source, and the rigidity is needed to account for
deflection in the Galactic magnetic field to backtrack to the source
direction.  To determine the distance and rigidity distributions we
simulate the propagation of a UHECR which originates as an iron
nucleus, through the cosmic microwave background radiation and the
extragalactic background light~\citep{Gilmore:2011ks} with {\scshape
  CRPropa3}~\citep{AlvesBatista:2016vpy}. We generated events
uniformly in distance and with an energy spectrum $\varpropto E^{-1}$
-- the approximate spectral index of UHECRs at the source, as deduced
from the combined spectrum and composition data.

The left panel of Fig.~\ref{fig:rig} shows a scatter plot of the
source distance and initial rigidity (= initial energy/26), of events
arriving at Earth with an energy of $E_{\atEarth}= E_\text{low}$ (the
corresponding plots for $E_\text{nom}$ look similar, but with a
reduced range in distance).  The opacity of the points is proportional
to $w_E \equiv \smash[b]{\exp[-\frac{1}{2}((E_{\atEarth} -
    \widehat{E})/\hat{\sigma})^2]}$, i.e.\ proportional to the
Gaussian probability to reconstruct $\widehat{E}$ given the simulated
energy $E_{\atEarth}$ at Earth, and the measurement uncertainty $
\hat{\sigma}$.  The right panel of Fig.~\ref{fig:rig} is a projection
in source distance of the left panel, displaying the histogram of
rigidities for each mass group.

\begin{figure}[t]
  \centering
  \includegraphics[clip,rviewport=0 0 1 1.1, width=0.8\linewidth]{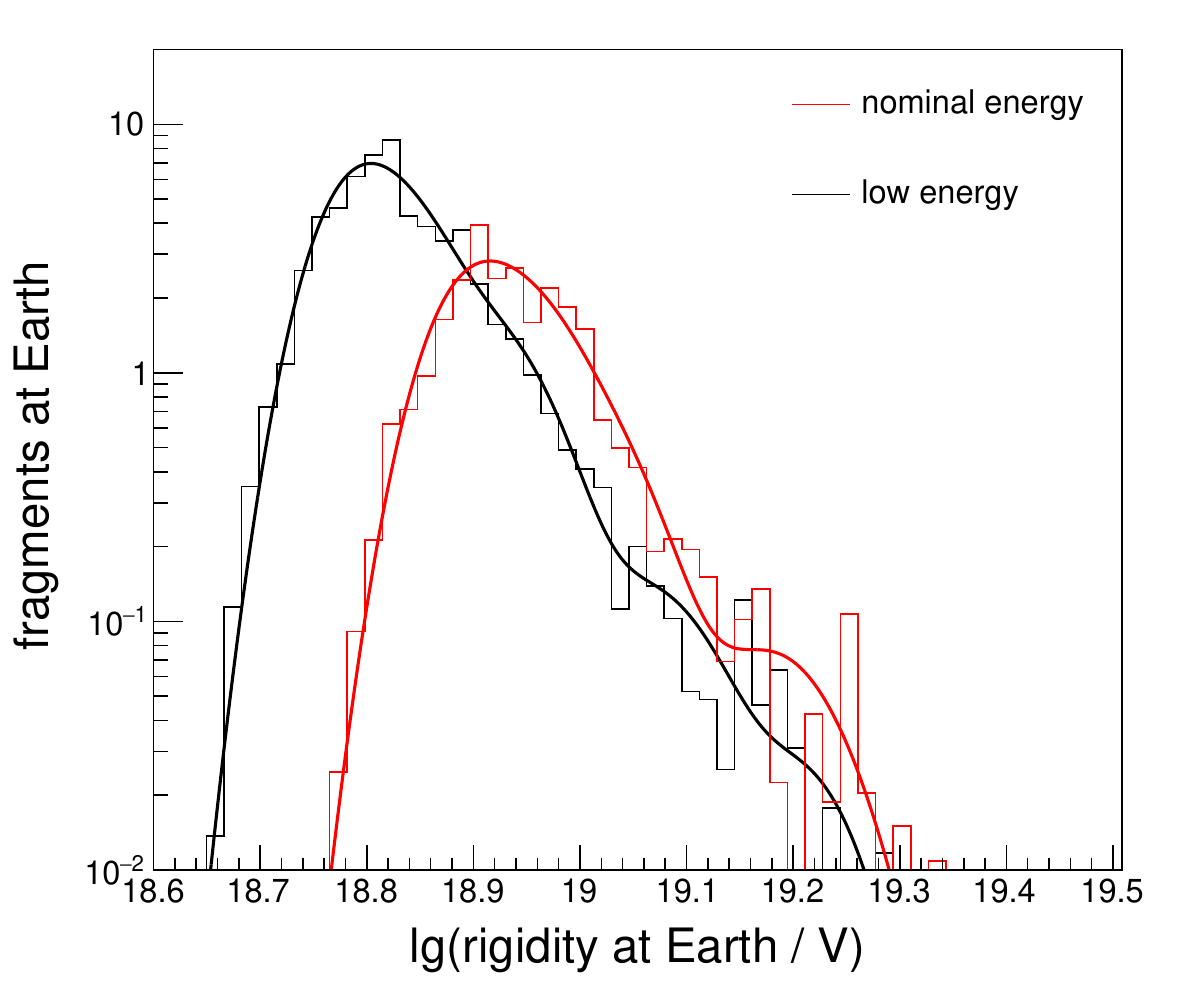}
  \caption{Distribution of the event rigidity at Earth, compatible with the
    detected energy at the nominal and low energy scales. \label{fig:rigDist}}
\end{figure}

Including a second weight $w_\R \equiv \exp(-\R/\R_\text{max})$ to
model the injected rigidity distribution, mostly suppresses the tail
of low-mass fragments, given a specified observed energy.  The fit of
TA data with iron primaries in~\citep{Unger:2015laa} suggests
$\R_\text{max} = 10^{18.7}$~V for the nominal energy scale; for the
low energy scale we use $\R_\text{max} = 10^{18.6}$~V.  Summing over
species, Fig.~\ref{fig:rigDist} shows the distribution of the
logarithm of the rigidity of the observed events, weighted with
$w_E\cdot w_\R$, for the two energy scales. The fraction by charge
group ($7<Z\leq 10$, $10<Z\leq 15$, $15<Z\leq 20$, $20<Z\leq 25$, Fe)
of detected UHECR, for iron injected at the source with a maximum
rigidity of $\lg \R_\text{max}/\text{V} = 18.7$ and 18.6, compatible
with the Amaterasu particle at the nominal and low energy scale is (0,
2.4, 13.4, 50.7, 33.4)\% and (0.4, 2.4, 15.6, 48.1, 33.3)\%,
respectively. Their mean and standard deviations are
\begin{equation}
  \lg(\R/\text{V}) = 18.94 {\pm} 0.07 \quad \text{and} \quad 18.83 {\pm} 0.07,
  \label{eq:rigdist}
\end{equation}
for the nominal and low energy scales, respectively.

\section{Propagation Distance}
\label{sec:dist}

\begin{figure}[t]
  \includegraphics[width=\linewidth]{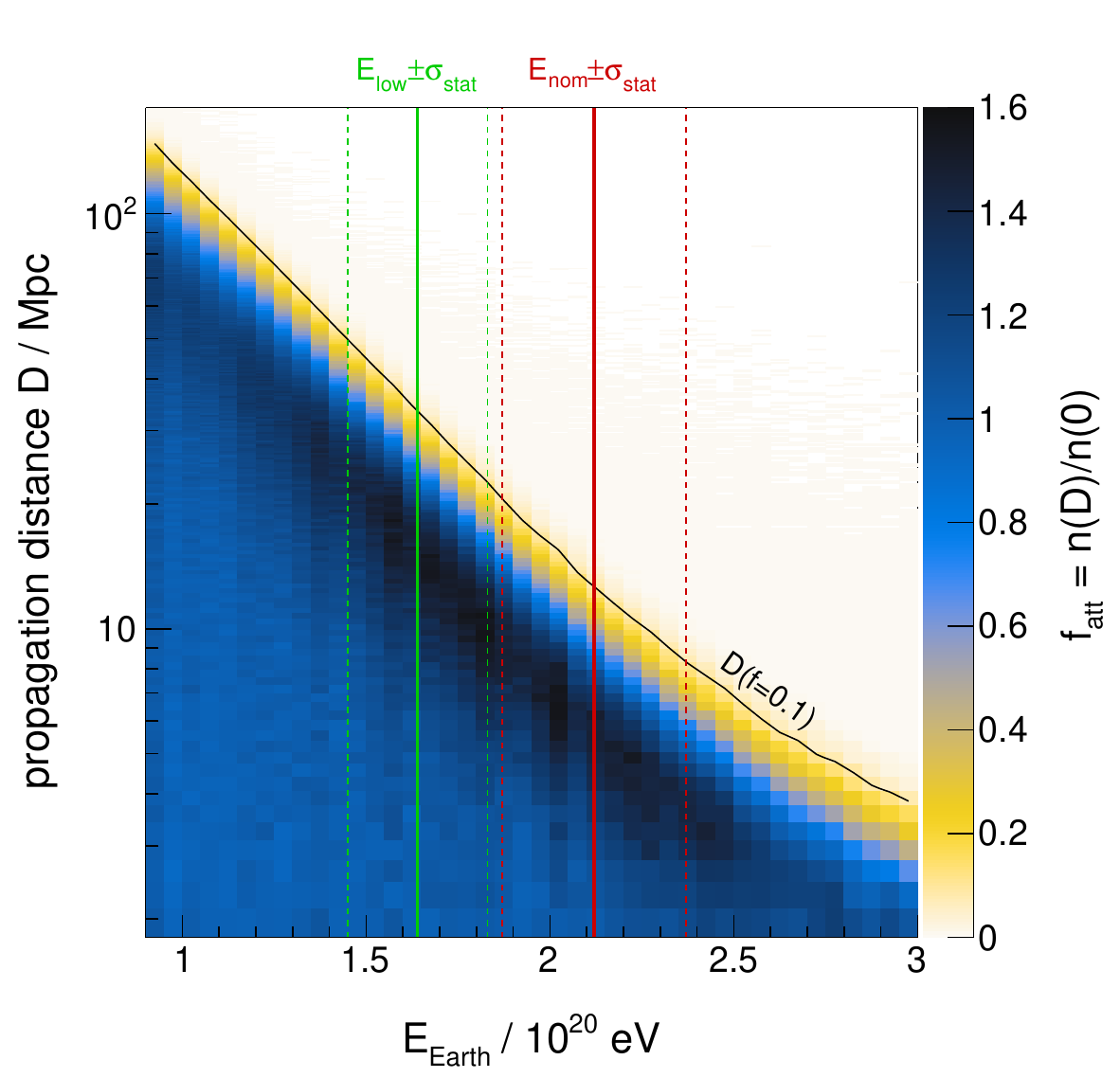}
  \caption{
     Propagation distances of CRs arriving at Earth,
    in each bin of energy as given on the $x$-axis of the plot, for iron injected with an $\R_\text{max} = 10^{18.6}$~V energy spectrum. The
    normalized number of CRs per distance bin is indicated by
    the color scale.  The nominal and low energy (thick central lines)
    of the Amaterasu particle, with the one-standard-deviation of the
    statistical reconstruction uncertainty (dashed lines), are
    superimposed. The distance, $D_{0.1}$, at which the relative number of
    arriving fragments drops below 10\% of the value at zero distance
    is shown as a black line.\label{fig:distDist}}
\end{figure}

The discussion of the previous section was for a fixed observed energy
value, $E_\text{nom}$ or $E_\text{low}$, but the experimental
statistical uncertainty needs to be taken into account.
Figure~\ref{fig:distDist} shows a histogram of the propagation
distance as a function of the UHECR energy at Earth, for $\R_\text{max} = 10^{18.6}$~V.  The vertical
lines mark the central and ${\pm} 1 \sigma_\text{stat}$ energy values.
One sees that the energy scale uncertainty is presently so great as to
lead to a factor-2.5 uncertainty on the source distance.

We define the approximate edge of the source volume as the distance,
$D_{0.1}$, at which the flux is attenuated by a factor-10 relative to
the case with no photo-interaction energy losses.  This distance is
$D_{0.1} =12.6^{+7.9}_{-4.3}$~Mpc and $33.5_{-10.9}^{+16.3}$~Mpc for
the nominal and low energy scales, respectively.  The uncertainties
originate from the one-sigma statistical uncertainty on the particle
energy; there is negligble sensitivity to the $\R_\text{max}$ choice.  Thus, taking the lower and upper 1-$\sigma$ bounds of these
uncertainties, the outer radius of the volume of possible sources is between
8 and 50~Mpc, for a factor of 240 uncertainty in the volume containing
the source.\footnote{Note that our definition of the horizon
  differs from that used in \citet{Kuznetsov:2023jfw}, where the horizon is defined to be
  the distance within which 95\% of the integral flux above a specified energy originated, assuming a uniform source distribution and an $E^{-1.89}$ spectrum without cutoff.
  Here we are interested in the probability distribution in distance
  of the individual source responsible for this particular event, therefore
  in this context our definition is more suitable.
  }

\begin{figure*}[t!]
  \centering
  \def\figh{0.18}
  \subfigure[][$\lg(\R/\text{V}) = 18.94$, $B$\&$b$, $n_b=1$]{
  \includegraphics[clip,rviewport=0 0 0.85 1,height=\figh\textheight]{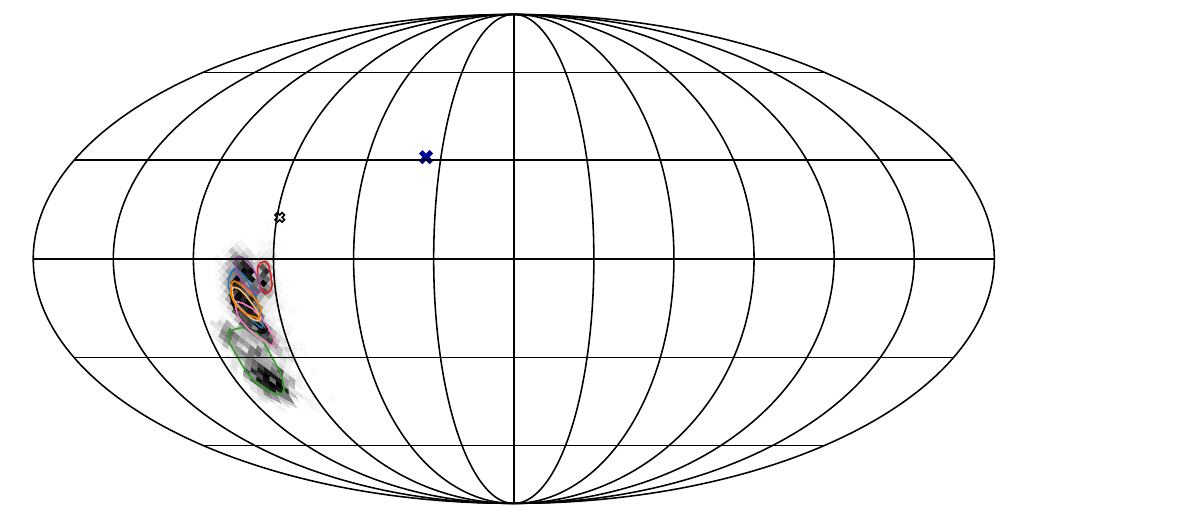}
  \label{fig:map0}
  }%
  \subfigure[][$\lg(\R/\text{V}) = 18.83$, $B$\&$b$, $n_b=1$]{
  \includegraphics[height=\figh\textheight]{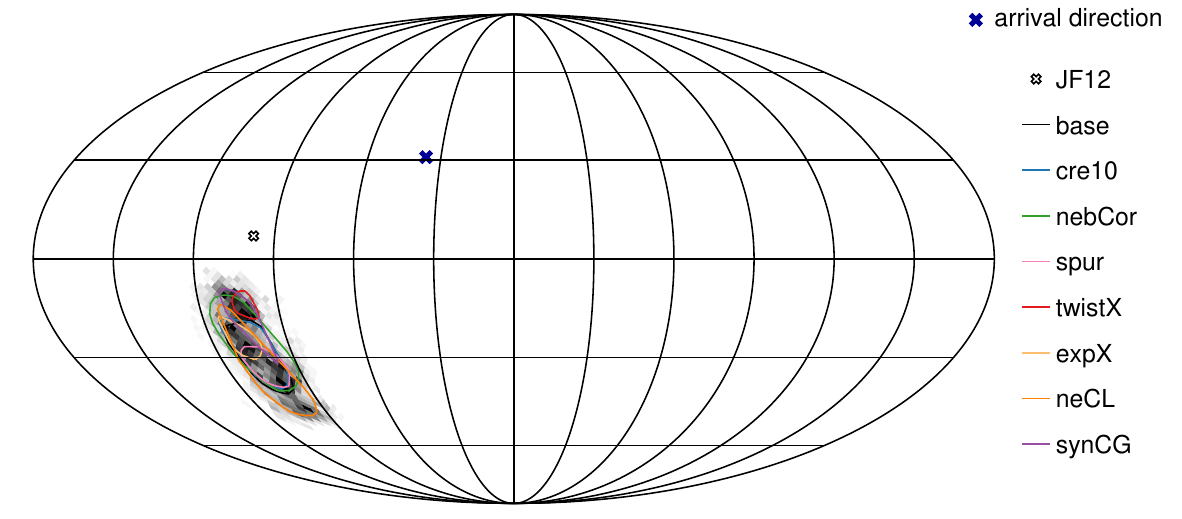}
  \label{fig:map1}
  }\\
  \subfigure[][$\lg(\R/\text{V}) = 18.94$, $B$\&$b$, $n_b=35$]{
  \includegraphics[clip,rviewport=0 0 0.85 1,height=\figh\textheight]{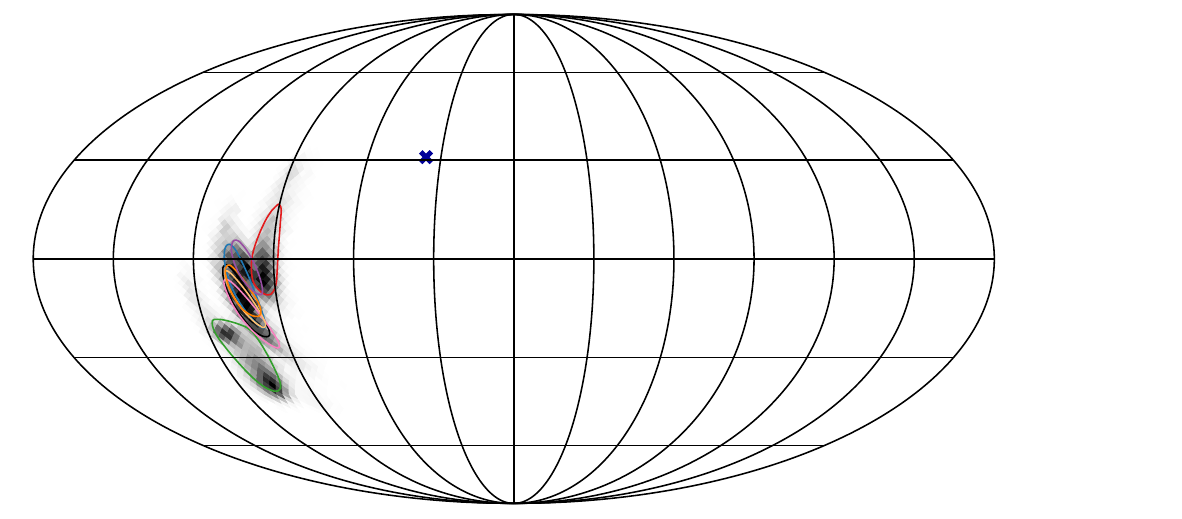}
  \label{fig:map2}
  }%
  \subfigure[][$\lg(\R/\text{V}) = 18.83$, $B$\&$b$, $n_b=35$]{
  \includegraphics[height=\figh\textheight]{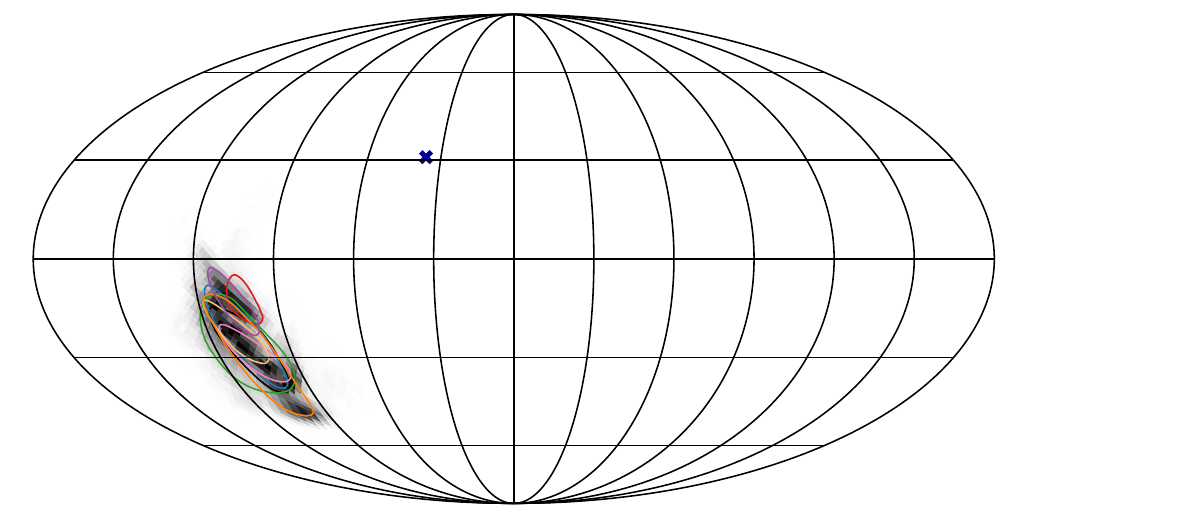}
  \label{fig:map3}
  }\\
  \subfigure[][$f(\R)_\text{nom}$, $B$\&$b$, $n_b=35$]{
  \includegraphics[clip,rviewport=0 0 0.85 1,height=\figh\textheight]{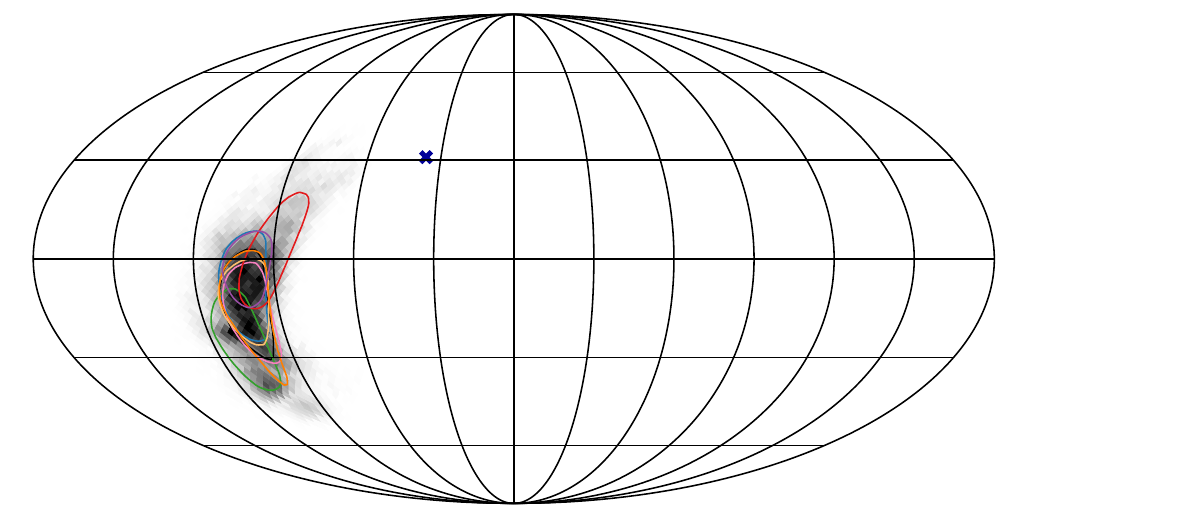}
  \label{fig:map4}
  }%
  \subfigure[][$f(\R)_\text{low}$, $B$\&$b$, $n_b=35$]{
  \includegraphics[height=\figh\textheight]{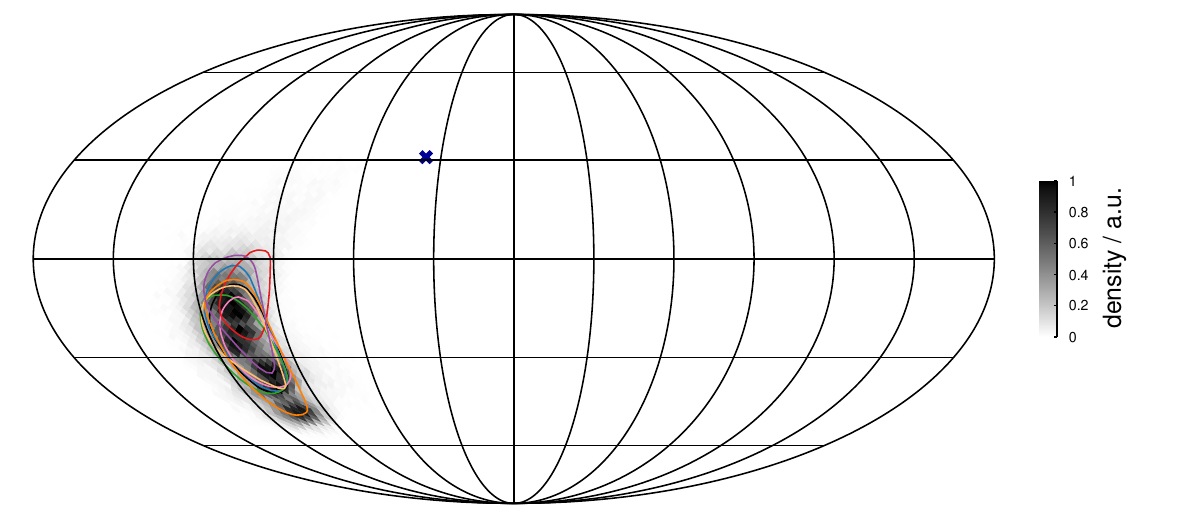}
  \label{fig:map5}
  }%
  \caption{Density of arrival directions back-tracked to the edge of
    the Galaxy for eight GMF model variations from
    \citet{Unger:2023lob}; results for the nominal and low energy
    scales are shown in the left and right columns, respectively. We
    use Galactic coordinates with the Galactic center at the origin
    and the longitude increasing towards the left. The measured
    arrival direction of the Amaterasu particle is shown as a blue
    cross. For reference, the back-tracked direction in the coherent
    field of \citet{Jansson:2012pc} is shown as an open cross in the
    top row.  For all panels, the magnetic field is a superposition of
    a regular ($B$) and a turbulent ($b$) component.  The top row
    shows one particular realization of the turbulent field ($n_b=1$).
    The middle and bottom rows show a superposition of the results for
    35 different realizations of $b$ ($n_b=35$).  The top and middle
    panels were calculated at a fixed rigidity (the mean values given
    in Eq.~(\ref{eq:rigdist})) while the bottom row is for the full
    distribution of rigidities, cf. Fig.~\ref{fig:rigDist}.  The
    colored contours show the 68\% confidence level convex hulls for each GMF
    model. \label{fig:backtrack}}
\end{figure*}

\section{Arrival Direction}
\label{sec:arrdir}
 We backtracked Amaterasu through the Galactic magnetic field to
identify the domain of highest source likelihood, using the ensemble
of eight models of the coherent Galactic magnetic field from
\citet{Unger:2023lob} that are compatible with the existing
astrophysical tracers of the magnetic field of the Milky Way and
encompass the range of uncertainty in the GMF.  We generate realizations of the
random field with the power spectrum taken to be that of a
turbulent cascade~\citep{1941DoSSR..30..301K} using the method
of~\citet{1999ApJ...520..204G}. We adopt 100~pc for the outer scale of
the turbulence as expected for supernova-driven turbulence and produce
a unit-norm random field from a superposition of 300 random waves
spaced logarithmically between 0.1 and 100~pc, corresponding to a
coherence length of 20~pc.  This unit-norm random field is then
weighted by the root-mean-square field strength of the Planck-tune of
the JF12 random field~\citep{Jansson:2012rt,Planck:2016gdp}.

Figure~\ref{fig:backtrack} shows, in Galactic coordinates, the
ensemble of backtracked events for the nominal (left column) and low
energy (right column) median rigidity assignments, for 1000 events
with a $\approx$0.5$^\circ$ Gaussian uncertainty in the measured
$(\text{RA},\text{Dec})$ angles of the arrival direction.  The first
row shows the result for a unique rigidity and single realization of
the random field.  The next row shows the superposition of results for
35 different realizations of the random field.  Finally, the bottom
row shows the source direction distribution summing over the
uncertainty in the rigidity distribution shown in
Fig.~\ref{fig:rigDist}.  In each case, the colored lines show the 68\%
convex hull contours~\citep{convexHull} of the backtracking results
for each of the eight GMF model variations.  The open crosses in the
top row show the back-tracked central arrival directions for the JF12
coherent field.

The meaning of the gray scale is somewhat nontrivial because we do not
know a priori which of the eight coherent field models of
\citet{Unger:2023lob} is closest to the truth.  An arrival direction
should not be deemed unlikely because it is far from the bulk of the
models if it is compatible with the (unknown to us) best model.
Therefore we proceed as follows.  The density of back-tracked
particles is displayed with gray scales on an $N_\text{side}=32$
{\scshape HEALPix} grid. Here we first collect the density maps of
each model $j$ separately and normalize the density in each pixel $i$
to the maximum density value for the given map, $\rho_{ij} = N_{ij} / N_{\text{max}, j}$. Here
$N_{ij}$ is the number of backtracked particles in pixel $i$ of the
skymap of GMF model $j$ and $N_{\text{max}, j} = \max_{1 \leq j \leq
  n} N_{ij} $ with the number of pixels $n$.  The union of the density
of all models is then
\begin{equation} \rho_{i} = \max_{1 \leq j \leq 8} \rho_{ij}.
  \label{eq:rho}
\end{equation}
I.e.,\ for a particular direction $i$ of the sky, $\rho_{i}$ is the
density of the closest GMF model variation.

The resulting source direction uncertainty is a compounded combination
of the uncertainty in arrival direction; the uncertainty in the
coherent field reflected in the model-to-model differences between the
eight different GMF models, which is amplified due to random fields on
top of the coherent one; the ``Galactic variance" from different
realizations of the random field; and the uncertainty in the rigidity
which is responsible for the difference between the middle and last
rows.  Note that the overall coherent deflection is larger than the
uncertainty in the deflections.

We define the source localization region to be the area on the sky in
which the density defined in Eq.~\eqref{eq:rho} is larger than
0.05. For both energy scales, this results in a solid angle of 3.5\%
of $4\uppi$ for the maps in the middle panel of
Fig.~\ref{fig:backtrack}, i.e.\ when including the uncertainties
originating from the magnetic field (coherent models, random field and
Galactic variance). Including the rigidity distribution (lower panel
of Fig.~\ref{fig:backtrack}) yields a larger value of 5.5\% and 4.6\%
for the nominal and low energy scales, respectively. Finally, the full
uncertainty, including the energy of Amaterasu, is conservatively
estimated by using the maximum of the two density maps of
Figs.~\ref{fig:map4} and \ref{fig:map5}. This yields a localization
uncertainty $\Omega_\text{loc}$ of 6.6\% of $4\uppi$ or 2726
deg$^2$. About half of this uncertainty can be attributed to the GMF
and the other half originates from the energy uncertainty (statistical
and systematic)\footnote{The typical standard deviation of the
  deflection angle after traversing a distance $D$ the turbulent
  extragalactic magnetic field (EGMF) of strength $B$ with coherence
  length $\lambda$ is $\smash[b]{\theta_\text{eg} = 2.5^\circ \,
    \sqrt{D/l} \, (\lambda/1\,\text{Mpc})\, (B/1\,\text{nG}) /
    (\R/10^{19}~\text{V})}$~\citep{1996ApJ...472L..89W}. The
  contribution to the estimated localization uncertainty can be
  neglected if $\theta_\text{eg} < \theta_\text{loc}$ where
  $\theta_\text{loc} = \arccos(1-\Omega_\text{loc}/2\pi) =
  28^\circ$. Assuming a conservative value of $\lambda =
  1\,\text{Mpc}$ and the upper limit on the EGMF of $B \leq
  1~nG$~\citep{Durrer:2013pga,2016PhRvL.116s1302P} yields
  $\theta_\text{eg} \leq 21^\circ$ and $\theta_\text{eg} \leq
  10^\circ$ for the propagation horizon $D_{0.1}$ and rigidity
  corresponding to $E_\text{low}$ and $E_\text{nom}$,
  respectively. The total localization angle $\smash[b]{\theta_\text{tot} =
  \sqrt{\theta_\text{loc}^2 + \theta_\text{eg}^2}}$ is thus at most
  25\% larger than $\theta_\text{loc}$, i.e.\ the localization is
  robust concerning the EGMF.}.

\section{Comparison to Galaxy Catalogues}
\label{sec:cata}

As a preliminary step to identifying astrophysical objects that could
be responsible for accelerating Amaterasu, we identify objects in
three catalogues used by the Pierre Auger
Collaboration~\citep{PierreAuger:2022qcg} that fall within the
localization volume.  These catalogs are a) the flux-limited Two
Micron All-Sky galaxy survey~\citep{Huchra:2011ii} cross-matched with
the HyperLEDA distance database~\citep{Makarov:2014txa} (``2MASS"
below), b) the flux-limited \textit{Swift}-BAT 105-month catalog of
active galactic nuclei (AGN) observed in hard X-rays
\citep{2018ApJS..235....4O} and c) a sample of nearby starburst
galaxies from \citet{2019JCAP...10..073L}.  In addition, we
investigate d) radio galaxies from the volume-limited catalogue
of~\citet{vanVelzen:2012fn}. Not all radio galaxies pass the
luminosity criterion~\citep{Waxman:1995vg,
  Blandford:1999hi,Farrar:2008ex} to be viable candidates for
accelerators up to the rigidity of the Amaterasu particle and we
therefore use only the subset identified as satisfying that
requirement by \citet{Matthews:2018laz}.  For completeness, we also
included galaxies from the "local volume catalogue" of
\cite{2018MNRAS.479.4136K}, a volume-limited sample of galaxies up to
a distance of 11~Mpc.

\begin{figure*}[t]
  \def\figw{0.35}
  \centering
  \subfigure[][Distribution of galaxies up to $D=125~\text{Mpc}$. The area zoomed in the lower panels (b)-(e) is indicated by a black rectangle.]{
  \includegraphics[clip,rviewport=0 0 1.05 1,width=0.8\linewidth]{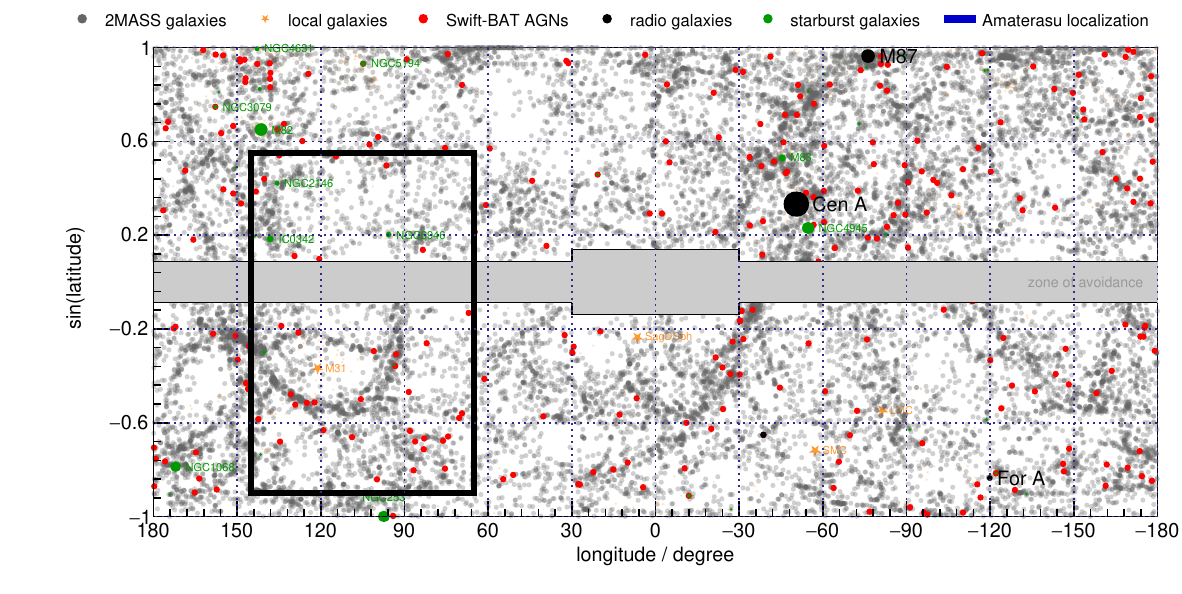}
  \label{fig:zoom150}
  }\\
  \subfigure[][$E_\text{low} - 2\,\sigma$, $D_{0.1}=72~\text{Mpc}$]{
  \includegraphics[clip, rviewport=0 0 1 0.93,width=\figw\linewidth]{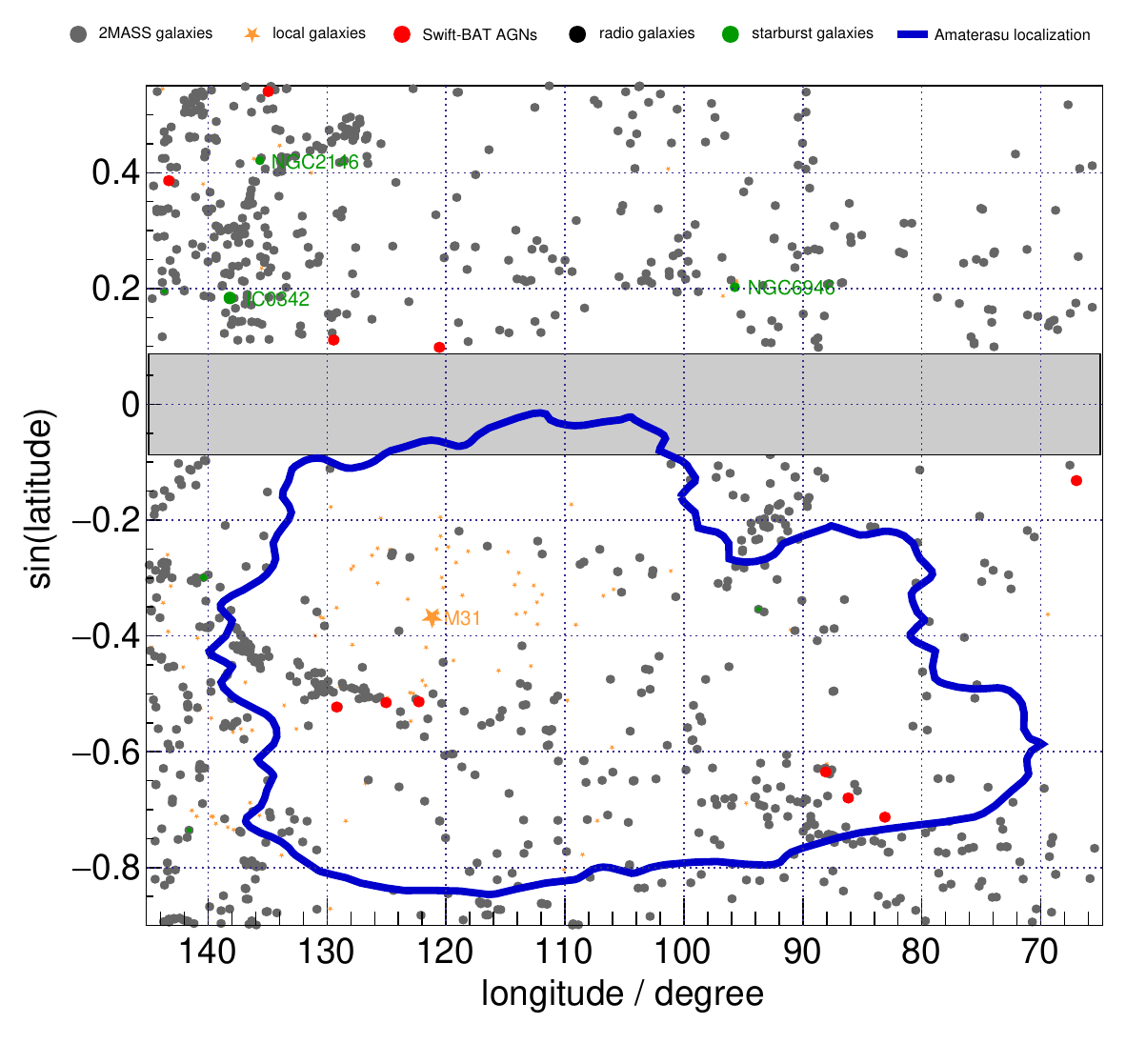}
  \label{fig:zoom5}
  }
  \subfigure[][$E_\text{low} - 1\,\sigma$, $D_{0.1}=42$~Mpc]{
  \includegraphics[clip, rviewport=0 0 1 0.93,width=\figw\linewidth]{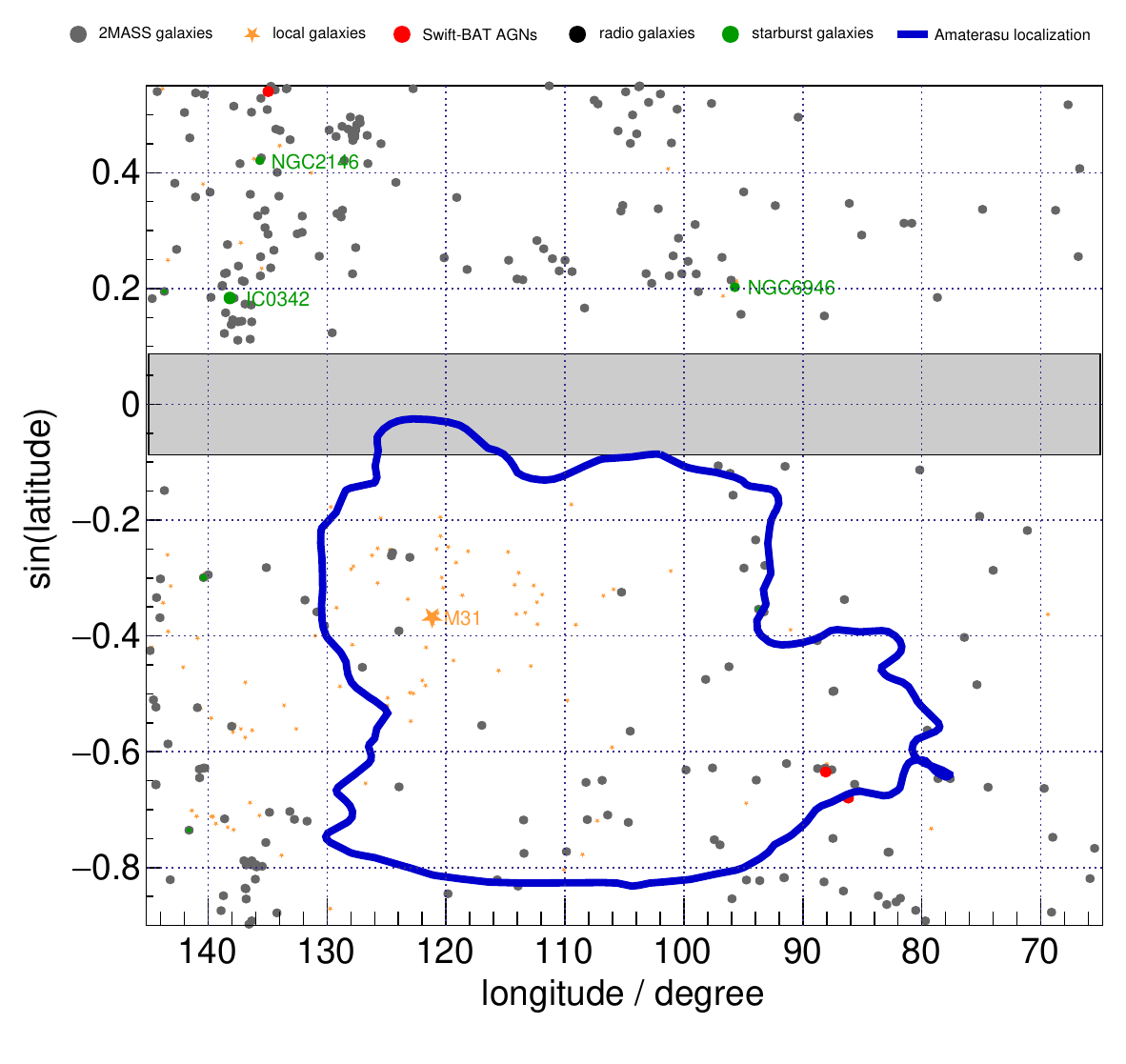}
  \label{fig:zoom4}
  }\\
  \subfigure[][$E_\text{low}$, $D_{0.1}=25$~Mpc]{
  \includegraphics[clip, rviewport=0 0 1 0.93,width=\figw\linewidth]{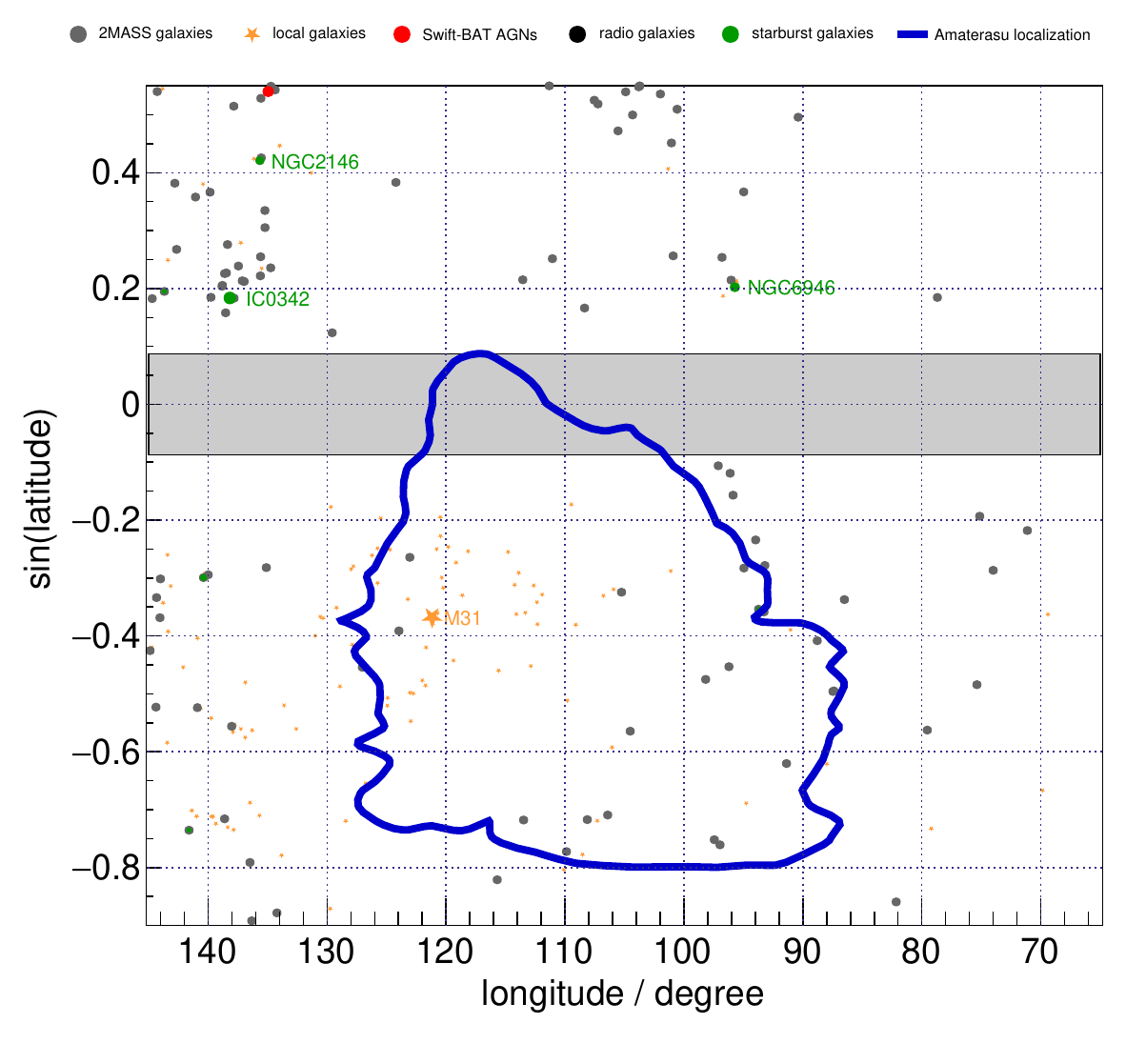}
  \label{fig:zoom1}
  }
  \subfigure[][$E_\text{nom}$, $D_{0.1}=10$~Mpc]{
  \includegraphics[clip, rviewport=0 0 1 0.93,width=\figw\linewidth]{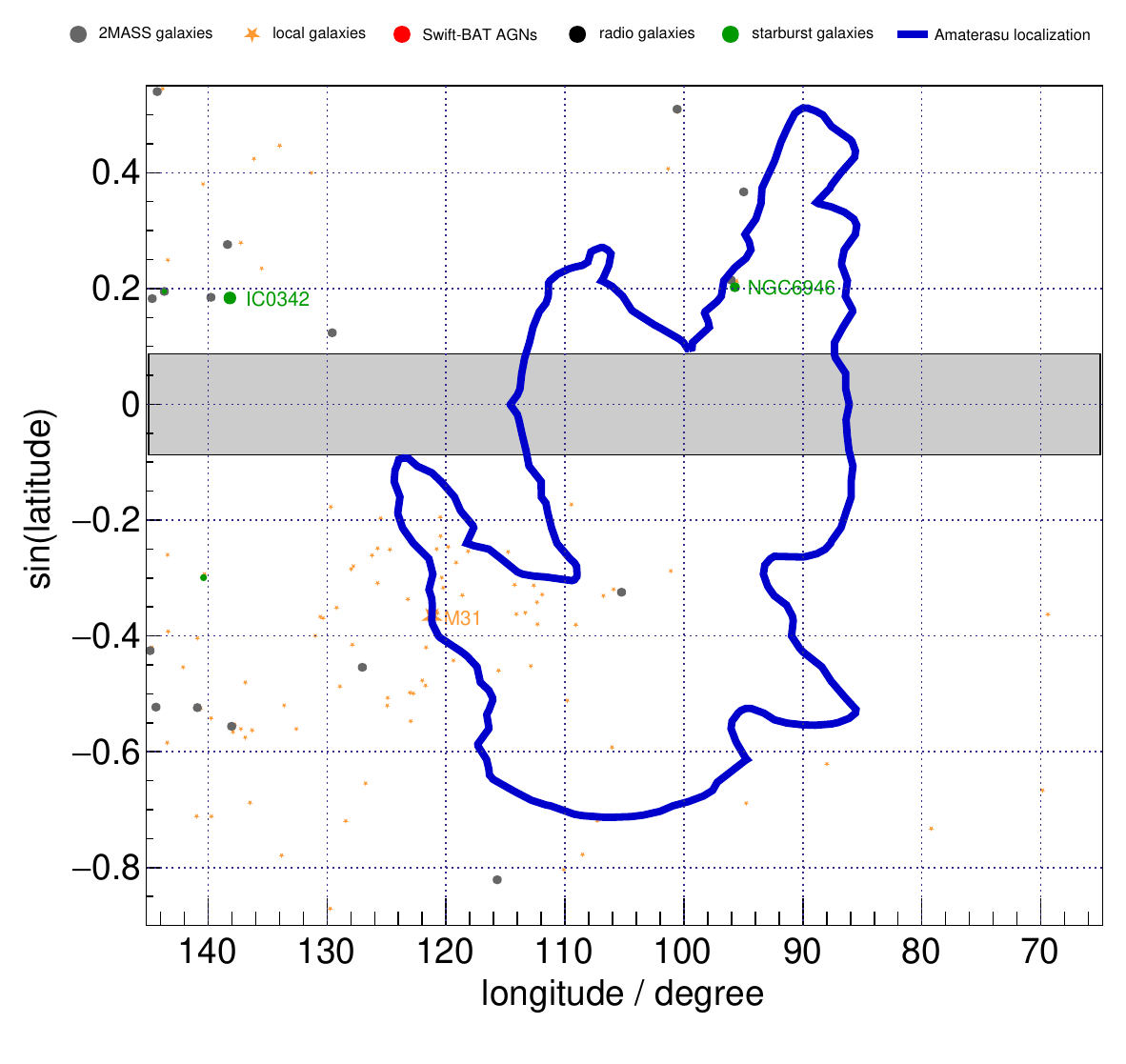}
  \label{fig:zoom0}
  }
  \caption{Galaxies within various distances for comparison to the
    source localization domain in Galactic coordinates. The top panel
    shows all galaxies up to a distance of $150$~Mpc, the other plots
    include galaxies up to the attenuation horizon, $D_{0.1}$
    depending on the rigidities corresponding to the particle energy
    assumed, see Eq.\ref{eq:rigdist} particle energy
    indicated. Contours of the distribution of backtracked particles
    are shown in blue at $\rho = 0.05$, see Eq.~(\ref{eq:rho}). The
    different colors denote galaxies from the 2MASS survey (gray),
    galaxies in the local volume, $D<11$ Mpc (orange), AGN (red),
    radio galaxies (black) and starburst galaxies
    (green).} \label{fig:catalogues}
\end{figure*}

The position of each of these galaxies within a distance of
$D=125~\text{Mpc}$ is displayed in Fig.~\ref{fig:zoom150}. The
galaxies from the 2MASS and \textit{Swift}-BAT catalogues are shown as
gray and red points, respectively.  Radio galaxies (RGs) and starburst
galaxies (SBGs) with flux greater than 5\% that of the brightest local
source of their type, i.e.,\ Cen~A (RGs) and M82 (SBGs), are shown
with black and green symbols whose area is proportional to the 1.1 and
1.4 GHz radio flux, respectively.  (The 1.4 GHz radio flux was taken
as a proxy for UHECR production by~\citet{PierreAuger:2022qcg}.)
Galaxies with a flux greater than 10\% the brightest local source
are labeled with names.  Galaxies from the Local Volume Catalog are
shown in orange and the four galaxies that dominate the local volume
in terms of stellar mass divided by distance-squared are shown with
labels and large symbols. The subsequent panels zoom into the
localization region of the Amaterasu source, starting $2
\sigma_\text{stat}$ below $E_\text{low}$ and increasing to
$E_\text{nom}$, showing the correspondingly shrinking horizon as a
function of the actual energy of the event.

We thus have identified, for each Amaterasu energy assignment, a very
conservative (maximal) source volume -- the angular locus and maximum
depth of the source -- taking into account the uncertainty in GMF and
rigidity. For each energy assignment, we can ask what candidate
sources fall within the volume.  A source class that is entirely
absent from the volume cannot be responsible for Amaterasu, at that
energy.  If there are some candidates, then we assess the plausibility
of that class by comparing its flux proxy  for
the sources inside the source localization to the total flux proxy
within the horizon. Since we are discussing a single event, the flux
ratio gives the probability that, under the hypothesis that the UHECR
flux follows the proxy, the Amaterasu source happens to fall in that particular
region of the sky.

We start with radio galaxies, long a favorite source candidate.  No
radio galaxies satisfying the luminosity criteria lie within the
localization volume, unless Amaterasu's energy is at least $2
\sigma_{\text{syst.}}$ lower than $E_\text{low}$; then, 3 galaxies
appear (NGC0262, NGC0315 and NGC7626).  However, the three
galaxies contribute only 0.2\% of the total flux, from which
alone one could exclude an origin of Amaterasu in radio galaxies
at the 3-$\sigma$ level. Taking into account that in addition, we required
a 2-$\sigma$ downward fluctuation of $E_\text{low}$ to increase the horizon,
makes this source hypothesis very unlikely.

Ordinary Swift-BAT AGN are another candidate.  Six out of 130 of them
with $D\leq 72~\text{Mpc}$ are within the localization shown in
Fig.~\ref{fig:zoom5} for energy $E_\text{low}-2\sigma_\text{stat.}$,
but none for a higher energy assignment. These 6 galaxies contribute
1\% to the total X-ray flux from \textit{Swift}-BAT galaxies within
the 72 Mpc horizon at this energy assignment.
Similarly to the case of radio galaxies, the combination of having to go to such a low energy
assignment in order to have any Swift-BAT AGN candidates, and the low
combined flux of the candidate sources relative to those within the
horizon but in other directions, is in strong tension with x-ray AGNs
being the source of Amaterasu.

The Amaterasu event is also in tension with the proposal that UHECR
sources are found in starburst galaxies.  No starburst galaxy falls
within the source direction domain for the three low energy
assignments.  Only at the nominal energy is the angular deflection
small enough that the starburst NGC6946 lies just at the edge of the
angular localization region using the ``twistX'' GMF-variant (c.f.,
Fig.~\ref{fig:map4}) as can be seen in Fig.~\ref{fig:zoom0}. (We
recall that at the boundary, the weight in the source direction
distribution is 0.05 times its peak value for the most favorable GMF
model, so some directional tension remains even with the twistX GMF
model.)  The 1.4 GHz radio flux of this galaxy, used as UHECR flux
proxy in \citet{PierreAuger:2018qvk} and \citet{PierreAuger:2022qcg},
is about 20\% of the flux of the brightest SBGs of the catalogue, M82
and NGC4945 (a candidate for causing the ``Auger hotspot''), and
contributes 3\% to the summed flux of all sources within the horizon
at this energy, $D_{0.1}=10$~Mpc.

For reference, the contribution of the K-band flux of 2MASS galaxies
within the localization volume, is at the 1-2\% level relative to the
all-sky flux of 2MASS galaxies within the horizon, for all four
energies displayed.  Given that the localization area is of order
6.6\% of the sky, the relative paucity of flux reflects the general
under-density of galaxies in the source volume, already noted by TA.
As a matter of information, but only relevant for the nominal energy
scenario, the under-density disappears when using galaxies in the
local volume catalog.  In this region the source volume ``flux''
(stellar mass per distance-squared) is dominated by M31 (Andromeda)
and is 6\% of the total.

\section{Summary and Conclusions}
\label{sec:concl}
We have studied the likely composition and origin of the UHECR event
Amaterasu, recently reported by the Telescope Array
collaboration~\citep{TAScience}.  We find that Amaterasu fits nicely
into the existing accumulated understanding of UHECR composition and
spectrum, if it is an iron nucleus or fragment thereof.  By contrast,
identifying Amaterasu as a proton or light nucleus would demand that
it be produced by an entirely new source class.  Adopting, therefore,
the minimalist interpretation of an intermediate or heavy nucleus and
following the prescription for the corresponding energy assignment
given by~\citet{TAScience}, we considered two energies for the
particle: the nominal energy of $E_\text{nom} = (2.12{\pm}0.25) \times
10^{20}$~eV and $1 \sigma_{\text{syst.}}$ lower $E_\text{low} =
(1.64{\pm}0.19) \times 10^{20}$~eV.  Detecting an event in this energy
range is natural -- even expected -- given accumulated exposure of TA,
based on extrapolating the spectrum already reported by TA; see Fig.~4
of \citet{Kim:2023eul}.

Taking into account interactions with the extragalactic background
light en route to Earth, and assuming that Amaterasu originated as an
iron nucleus, we conclude it was produced by a source whose distance
is within 8~Mpc, or up to 50~Mpc, depending on its energy.  About one
third of the time the original nucleus survives intact and about half
the time its charge upon reaching Earth is in the range of 20-25; only
a few percent of cases arrive with $Z\leq 15$.

We backtracked Amaterasu through the Galactic magnetic field to
identify the domain of highest source likelihood.  Altogether we used
$8 \times 35$ different GMF realizations, based on the 8 coherent GMF
models in the UF23 suite~\citep{Unger:2023lob}, which in their
ensemble encompass the range of uncertainty in the large scale
Galactic field. Added to each of the coherent GMF models was one of 35
different random field realizations, whose field strength is given by
the Planck-tune of the JF12 random field~\citep{Planck:2016gdp}.
Taking the union of field models and Amaterasu rigidities, we obtain a
conservative angular locus for the source, for each energy assignment.
Combined with the source distance constraint, we identify the
most-probable source volume for each Amaterasu energy assignment.

Finally, interrogating catalogs of sufficiently-powerful radio
galaxies, Swift-BAT AGNs, starburst galaxies and generic galaxies, we
assess which of these galaxy types can have produced Amaterasu.  We
find that none of the ``usual suspects" among candidate non-transient
UHECR accelerators provides a comfortable explanation.  There are
simply no radio galaxies of sufficient power within the localization
volume, unless the energy is $2 \sigma_\text{stat.}$ below
$E_\text{low}$.  Moreover even though the horizon is so large in this
case that a few radio galaxy candidates appear in the source volume,
their total UHECR flux proxy
is only a fraction of a percent of that of the rest of the sky.
Similarly, only for this lowest energy
assignment are there any candidate Swift-BAT AGNs, and those are also
deprecated by having a small total flux-proxy relative to Swift-BAT
AGNs in the rest of the sky.

A third popular candidate to host UHECR accelerators is starburst
galaxies.  SBGs are of primary interest because long GRBs result from
the collapse of massive young stars, hence their rate is enhanced in
SGBs.  However, SBGs are also deprecated as a possible source of
Amaterasu by this analysis.  Only for the highest energy assignment is
there a SBG (barely) in the localization domain, for a single GMF
model, and its flux proxy is only 5\% that of SBGs outside the
localization region.

The most straightforward explanation for Amaterasu seems to be that it
was produced by a transient in an otherwise ordinary galaxy.  Several
types of transients found in ordinary galaxies merit consideration,
including tidal disruption events~\citep{Farrar:2008ex}, young
magnetars~\citep{Blasi:2000xm,Arons:2002yj} and potentially jet formation in
neutron star-black hole (or possibly binary neutron star) mergers.  A
similar conclusion was reached by \citet{Fitoussi:2019dfe}, who
attribute the nominally 320 EeV ``Fly's Eye" event to an iron nucleus
from a stellar transient.\footnote{\citet{Fitoussi:2019dfe} ignored
  the coherent deflections of the Fly's Eye event, detected at
  $(\ell,b) = (163.4, 9.6)^\circ$.  We checked that approximation,
  finding an average back-tracked extragalactic arrival direction of
  $(165\pm 3, 14.7\pm 8.5)^\circ$, assuming a rigidity of $\R=7.3$~EV,
  where the uncertainties give the maximum range of the predictions
  from the eight GMF models of ~\cite{Unger:2023lob}.}

Our work underscores the power of a single well-measured, high energy
event, combined with the ability to measure or estimate the
composition on theoretical grounds, for restricting the possible
sources of individual high energy UHECRs.  Future analysis of the TA
energy calibration may enable Amaterasu's energy to be more accurately
bracketed, which would increase the value of this single event even
further.  For instance, if the energy assignment is reduced
sufficiently below the current nominal value, SBGs would be eliminated
as a possible source, while if the lower bound on the energy can be
shown to be clearly above $\approx 130$ EeV, radio galaxies and
Swift-BAT AGNs could be excluded.

\section*{Acknowledgments}
We would like to thank Luis Anchordoqui, Ralph Engel, Toshihiro Fujii,
Mikhail Kuznetsov, and Foteini Oikonomou for useful discussions.  The
research of GRF has been supported by the National Science Foundation
grant NSF-PHY-2013199.

\bibliography{refs}{}

\begin{thebibliography}{}
\expandafter\ifx\csname natexlab\endcsname\relax\def\natexlab#1{#1}\fi
\providecommand{\url}[1]{\href{#1}{#1}}
\providecommand{\dodoi}[1]{doi:~\href{http://doi.org/#1}{\nolinkurl{#1}}}
\providecommand{\doeprint}[1]{\href{http://ascl.net/#1}{\nolinkurl{http://ascl.net/#1}}}
\providecommand{\doarXiv}[1]{\href{https://arxiv.org/abs/#1}{\nolinkurl{https://arxiv.org/abs/#1}}}

\bibitem[{Aab {et~al.}(2017)}]{PierreAuger:2016use}
Aab, A., {et~al.} 2017, JCAP, 04, 038, \dodoi{10.1088/1475-7516/2017/04/038}

\bibitem[{Aab {et~al.}(2018)}]{PierreAuger:2018qvk}
---. 2018, Astrophys. J. Lett., 853, L29, \dodoi{10.3847/2041-8213/aaa66d}

\bibitem[{Abbasi {et~al.}(2024)}]{TAScience}
Abbasi, R., {et~al.} 2024, Science, 382, 903, \dodoi{10.1126/science.abo5095}

\bibitem[{Abdul~Halim {et~al.}(2023)}]{PierreAuger:2022qcg}
Abdul~Halim, A., {et~al.} 2023, Astrophys. J. Suppl., 264, 50,
  \dodoi{10.3847/1538-4365/aca537}

\bibitem[{Abreu {et~al.}(2022)}]{PierreAuger:2022axr}
Abreu, P., {et~al.} 2022, Astrophys. J., 935, 170,
  \dodoi{10.3847/1538-4357/ac7d4e}

\bibitem[{Adam {et~al.}(2016)}]{Planck:2016gdp}
Adam, R., {et~al.} 2016, Astron. Astrophys., 596, A103,
  \dodoi{10.1051/0004-6361/201528033}

\bibitem[{Alves~Batista {et~al.}(2016)}]{AlvesBatista:2016vpy}
Alves~Batista, R., {et~al.} 2016, {JCAP}, 05, 038,
  \dodoi{10.1088/1475-7516/2016/05/038}

\bibitem[{Arons(2003)}]{Arons:2002yj}
Arons, J. 2003, Astrophys. J., 589, 871, \dodoi{10.1086/374776}

\bibitem[{Bird {et~al.}(1995)}]{HIRES:1994ijd}
Bird, D.~J., {et~al.} 1995, Astrophys. J., 441, 144, \dodoi{10.1086/175344}

\bibitem[{Blandford(2000)}]{Blandford:1999hi}
Blandford, R.~D. 2000, Phys. Scripta T, 85, 191,
  \dodoi{10.1238/Physica.Topical.085a00191}

\bibitem[{Blasi {et~al.}(2000)Blasi, Epstein, \& Olinto}]{Blasi:2000xm}
Blasi, P., Epstein, R.~I., \& Olinto, A.~V. 2000, Astrophys. J. Lett., 533,
  L123, \dodoi{10.1086/312626}

\bibitem[{Durrer \& Neronov(2013)}]{Durrer:2013pga}
Durrer, R., \& Neronov, A. 2013, Astron. Astrophys. Rev., 21, 62,
  \dodoi{10.1007/s00159-013-0062-7}

\bibitem[{Ehlert {et~al.}(2023)Ehlert, van Vliet, Oikonomou, \&
  Winter}]{Ehlert:2023btz}
Ehlert, D., van Vliet, A., Oikonomou, F., \& Winter, W. 2023.
\newblock \doarXiv{2304.07321}

\bibitem[{Farrar \& Gruzinov(2009)}]{Farrar:2008ex}
Farrar, G.~R., \& Gruzinov, A. 2009, Astrophys. J., 693, 329,
  \dodoi{10.1088/0004-637X/693/1/329}

\bibitem[{Fitoussi {et~al.}(2020)Fitoussi, Medina-Tanco, \&
  D'Olivo}]{Fitoussi:2019dfe}
Fitoussi, T., Medina-Tanco, G., \& D'Olivo, J.-C. 2020, JCAP, 01, 042,
  \dodoi{10.1088/1475-7516/2020/01/042}

\bibitem[{{Giacalone} \& {Jokipii}(1999)}]{1999ApJ...520..204G}
{Giacalone}, J., \& {Jokipii}, J.~R. 1999, ApJ, 520, 204,
  \dodoi{10.1086/307452}

\bibitem[{Gilmore {et~al.}(2012)Gilmore, Somerville, Primack, \&
  Dominguez}]{Gilmore:2011ks}
Gilmore, R.~C., Somerville, R.~S., Primack, J.~R., \& Dominguez, A. 2012,
  MNRAS, 422, 3189, \dodoi{10.1111/j.1365-2966.2012.20841.x}

\bibitem[{Halim {et~al.}(2023)}]{PierreAuger:2022atd}
Halim, A.~A., {et~al.} 2023, {JCAP}, 05, 024,
  \dodoi{10.1088/1475-7516/2023/05/024}

\bibitem[{Huchra {et~al.}(2012)}]{Huchra:2011ii}
Huchra, J.~P., {et~al.} 2012, Astrophys. J. Suppl., 199, 26,
  \dodoi{10.1088/0067-0049/199/2/26}

\bibitem[{Jansson \& Farrar(2012{\natexlab{a}})}]{Jansson:2012pc}
Jansson, R., \& Farrar, G.~R. 2012{\natexlab{a}}, Astrophys. J., 757, 14,
  \dodoi{10.1088/0004-637X/757/1/14}

\bibitem[{Jansson \& Farrar(2012{\natexlab{b}})}]{Jansson:2012rt}
---. 2012{\natexlab{b}}, Astrophys. J. Lett., 761, L11,
  \dodoi{10.1088/2041-8205/761/1/L11}

\bibitem[{{Karachentsev} {et~al.}(2018){Karachentsev}, {Kaisina}, \&
  {Makarov}}]{2018MNRAS.479.4136K}
{Karachentsev}, I.~D., {Kaisina}, E.~I., \& {Makarov}, D.~I. 2018, \mnras, 479,
  4136, \dodoi{10.1093/mnras/sty1774}

\bibitem[{Kim {et~al.}(2023)}]{Kim:2023eul}
Kim, J., {et~al.} 2023, EPJ Web Conf., 283, 02005,
  \dodoi{10.1051/epjconf/202328302005}

\bibitem[{{Kolmogorov}(1941)}]{1941DoSSR..30..301K}
{Kolmogorov}, A. 1941, Akademiia Nauk SSSR Doklady, 30, 301

\bibitem[{Kuznetsov(2023)}]{Kuznetsov:2023jfw}
Kuznetsov, M.~Y. 2023.
\newblock \doarXiv{2311.14628}

\bibitem[{{Lunardini} {et~al.}(2019){Lunardini}, {Vance}, {Emig}, \&
  {Windhorst}}]{2019JCAP...10..073L}
{Lunardini}, C., {Vance}, G.~S., {Emig}, K.~L., \& {Windhorst}, R.~A. 2019,
  JCAP, 2019, 073, \dodoi{10.1088/1475-7516/2019/10/073}

\bibitem[{Makarov {et~al.}(2014)Makarov, Prugniel, Terekhova, Courtois, \&
  Vauglin}]{Makarov:2014txa}
Makarov, D., Prugniel, P., Terekhova, N., Courtois, H., \& Vauglin, I. 2014,
  Astron. Astrophys., 570, A13, \dodoi{10.1051/0004-6361/201423496}

\bibitem[{Matthews {et~al.}(2018)Matthews, Bell, Blundell, \&
  Araudo}]{Matthews:2018laz}
Matthews, J.~H., Bell, A.~R., Blundell, K.~M., \& Araudo, A.~T. 2018, Mon. Not.
  Roy. Astron. Soc., 479, L76, \dodoi{10.1093/mnrasl/sly099}

\bibitem[{McDermott \& Lin(2007)}]{convexHull}
McDermott, J.~P., \& Lin, D. K.~J. 2007, IIE Transactions, 39, 581,
  \dodoi{10.1080/07408170600899599}

\bibitem[{Muzio {et~al.}(2022)Muzio, Farrar, \& Unger}]{Muzio:2021zud}
Muzio, M.~S., Farrar, G.~R., \& Unger, M. 2022, Phys. Rev. D, 105, 023022,
  \dodoi{10.1103/PhysRevD.105.023022}

\bibitem[{Muzio {et~al.}(2019)Muzio, Unger, \& Farrar}]{Muzio:2019leu}
Muzio, M.~S., Unger, M., \& Farrar, G.~R. 2019, Phys. Rev. D, 100, 103008,
  \dodoi{10.1103/PhysRevD.100.103008}

\bibitem[{{Oh} {et~al.}(2018){Oh}, {Koss}, {Markwardt},
  {et~al.}}]{2018ApJS..235....4O}
{Oh}, K., {Koss}, M., {Markwardt}, C.~B., {et~al.} 2018, ApJS, 235, 4,
  \dodoi{10.3847/1538-4365/aaa7fd}

\bibitem[{{Peters}(1961)}]{Peters:1961}
{Peters}, B. 1961, Il Nuovo Cimento, 22, 800, \dodoi{10.1007/BF02783106}

\bibitem[{{Pshirkov} {et~al.}(2016){Pshirkov}, {Tinyakov}, \&
  {Urban}}]{2016PhRvL.116s1302P}
{Pshirkov}, M.~S., {Tinyakov}, P.~G., \& {Urban}, F.~R. 2016, \prl, 116,
  191302, \dodoi{10.1103/PhysRevLett.116.191302}

\bibitem[{Unger \& Farrar(2023)}]{Unger:2023lob}
Unger, M., \& Farrar, G. 2023.
\newblock \doarXiv{2311.12120}

\bibitem[{Unger {et~al.}(2015)Unger, Farrar, \& Anchordoqui}]{Unger:2015laa}
Unger, M., Farrar, G.~R., \& Anchordoqui, L.~A. 2015, Phys. Rev. D, 92, 123001,
  \dodoi{10.1103/PhysRevD.92.123001}

\bibitem[{van Velzen {et~al.}(2012)van Velzen, Falcke, Schellart,
  Nierstenhoefer, \& Kampert}]{vanVelzen:2012fn}
van Velzen, S., Falcke, H., Schellart, P., Nierstenhoefer, N., \& Kampert,
  K.-H. 2012, Astron. Astrophys., 544, A18, \dodoi{10.1051/0004-6361/201219389}

\bibitem[{Waxman(1995)}]{Waxman:1995vg}
Waxman, E. 1995, Phys. Rev. Lett., 75, 386, \dodoi{10.1103/PhysRevLett.75.386}

\bibitem[{{Waxman} \& {Miralda-Escude}(1996)}]{1996ApJ...472L..89W}
{Waxman}, E., \& {Miralda-Escude}, J. 1996, \apjl, 472, L89,
  \dodoi{10.1086/310367}

\end{thebibliography}
\bibliographystyle{aasjournal}

\end{document}